\documentclass[prd,twocolumn,amsmath,amssymb]{revtex4}
\begin{document}
\title{{\sf RESEARCH NEWS :}\\{\bf DIRECT OBSERVATION OF NEUTRINO 
OSCILLATIONS AT THE SUDBURY NEUTRINO OBSERVATORY}}
\author{B. Ananthanarayan}
\affiliation{Centre for Theoretical Studies, Indian Institute of Science, 
Bangalore 560 012, India}
\author{Ritesh K. Singh}
\affiliation{Centre for Theoretical Studies, Indian Institute of Science, 
Bangalore 560 012, India}
\begin{abstract}
\noindent
A brief description of the path-breaking evidence for the
observation of neutrino oscillations at the
Sudbury Neutrino Observatory is presented and 
the experimental principles and theory thereof are briefly discussed.
\end{abstract}
\maketitle
\noindent
\begin{small}
The Sudbury Neutrino Observatory (SNO), located in Canada, has recently
announced evidence for the direct observation of oscillation of neutrino
flavors~\cite{nc}.  The observations indicate that during the flight from
the interior of the sun to the earth, the neutrinos produced as electron
type neutrinos change their flavor.
Neutrinos are fundamental particles that come in
three types (flavors): electron, muon and tau denoted by $\nu_e,\, \nu_\mu$ and
$\nu_\tau$ respectively.  Electron type 
(anti-)neutrinos are produced, for instance,
in radioactive decays of certain heavy nuclei, and are also produced
copiously in stars during the process of nucleosynthesis.  
It may be recalled that a long standing problem, called the ``solar neutrino
problem'', associated with the shortfall in the number of neutrinos observed 
on the earth as compared to the number expected from calculations based on the
standard solar model and the standard model of elementary particle physics. 
The solar models
happen to be very stringently constrained and the flux of neutrinos produced
in the sun is known to practically no uncertainty, due to extreme sensitivity
of the flux to the values of the parameters going into the solar model.
As a result, it has been long believed that any solution to the problem must
come from the elementary particle physics sector, such as the oscillation
of flavors, or the possible interaction of the neutrinos with the solar
magnetic fields which might significantly scatter away the neutrinos that
were directed towards the earth. 
A flavor oscillation necessarily implies that the neutrinos are massive
particles and was first pointed out a long time ago by the Italian born
Soviet scientist Bruno Pontecorvo.  
Whereas there are constraints on the masses of the electron,
muon and tau type neutrinos ($m_{\nu_e}\leq 2.8 \ {\rm eV}/c^2$, 
$m_{\nu_\mu}\leq 170 \ {\rm keV}/c^2$ and $m_{\nu_\tau}\leq 18.2 \ {\rm MeV}/c^2$),
the oscillation can, in
principle probe differences of the squares of the masses and the
probability of conversion from one flavor into another depends on the path
length between production and observation, and on the kinetic energy of
the neutrino that is being observed.
The observation from SNO definitively shows that the electron type neutrino 
oscillates into an admixture of the other two flavors.  The SNO collaboration 
also reports a possible day-night effect~\cite{dn}, 
in that there seems to be some indication that
the probability of conversion of the electrons depends on the path length
which depends on whether the observation is made during the day or during
the night.\\\\
The experiment is based on the
principle of detecting interactions of neutrinos with matter in an ingenious
manner, which uses heavy water as the medium in which the interactions
take place and in which the reaction products are detected.  This differs
from earlier measurements which used normal water, at Kamiokande and the
super-Kamiokande experiments, and from those based on radiochemical measurements
which used interactions of the neutrinos with nuclei of chlorine in
carbon tetrachloride or with gallium nuclei.  For completeness, we note
that the solar neutrino problem was first established by the historic
chlorine experiment of Ray Davis at the Homestake Mine in South
Dakota, USA, which employs the following reaction to detect neutrinos
\begin{eqnarray*}
& \displaystyle ^{37}{\rm Cl} + \nu_e \to \ ^{37}{\rm Ar} + e^-. &
\end{eqnarray*}
The Gallium experiments use the reaction
\begin{eqnarray*}
& \displaystyle ^{71}{\rm Ga} + \nu_e \to \ ^{71}{\rm Ge} + e^-
\end{eqnarray*}
The neutrinos that are observed at SNO are produced in the core of the sun in
the following reaction,
\begin{eqnarray*}
& \displaystyle ^8{\rm B} \to \ ^8{\rm Be} + e^+ + \nu_e .&
\end{eqnarray*}
The kinetic energy of the neutrinos produced in this reaction could vary
from as little as a fraction of an MeV to approximately 15 MeV.
These neutrinos interact with the water molecules in a variety of ways.
The normal water detector detects this neutrino from its elastic scattering (ES)
with the electrons in the water molecules.  The ES measurements are
predominantly sensitive to the electron type flavor only.  The measurements of 
super-Kamiokande had established from the event rates they observed, a shortfall
in the expected event rate, consistent with the earlier radiochemical
measurements which had led to the classical solar neutrino problem.
These observations were based on the charged current (CC)
reactions.  The full electroweak theory of Glashow, Salam and Weinberg
has established that the theory also has, in addition to the conventional
electromagnetic interaction, what is called the neutral current 
(NC) interactions. All the interactions arise from the exchange of
(virtual) intermediate vector bosons, the CC interactions from $W^\pm$, 
the NC interactions arise from
a neutral boson called $Z^0$, just as the electromagnetic
interactions arise from the photon. 
 Here we note that the main principle in both
the normal and heavy water detectors for the observation of the scattered
electron in the ES or the produced electron in the CC reaction is that of
the detection of \v{C}erenkov light produced by the ultra-relativistic
electron during its motion with its velocity exceeding that of light
in the medium (water).   The \v{C}erenkov light is detected by photomultiplier
tubes at the boundary of containers.  Note also that the water has to be of
extraordinary purity in order to prevent attenuation of the light during
it travel from the electron source to the detector.
%
%
The measurements of the ES and CC reaction rates by the SNO Collaboration
were reported earlier~\cite{rate}, which further confirmed the 
solar neutrino problem.\\\\
The remarkable advantage of the 
heavy water detector is its capability to observe the NC 
interactions as well, in addition to observing the CC interactions.
It must be mentioned that the construction of the SNO experiment was
directly inspired by an important paper by the late Herb Chen~\cite{chen}.
The crucial property of the heavy water detector arises from the 
fact that the deuterium nucleus has a remarkably small binding energy
of 2.225 MeV. This may be contrasted with the typical binding energies of 8-9
MeV/nucleon for most nuclei. As a result, the kinetic energy of the neutrinos 
is sufficiently large so as to induce the following reactions
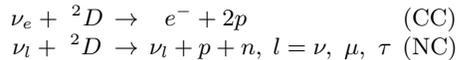
\begin{center}
\begin{tabular}{ccccc}
$ \nu_e + \ ^2D$ & $\to$ & $e^- + 2 p$     &                      &(CC)\\
$ \nu_l + \ ^2D$ & $\to$ & $\nu_l + p + n,$ &$l=\nu, \ \mu, \ \tau$&(NC) 
\end{tabular}
\end{center}
The reason why the CC reaction above is sensitive only to the electron-type
neutrino is that kinetic energy of the neutrinos produced in the boron
reaction is sufficient only to produce electrons ($m_e=0.511 \ {\rm MeV/c^2}$),
in accordance with Einstein's mass-energy equivalence,
whereas the muons and the tau-leptons are too massive to be produced
in this reaction ($m_\mu=105.7 \ {\rm MeV/c^2}, m_\tau=1777 {\rm \ MeV/c^2}$).
The NC reaction does not have this kinematic constraint and is therefore
sensitive to all flavors.
The combination of the small binding energy of the deuterium nucleus
with the NC reaction is capable of producing a characteristic signal which
can be detected when the deuterium nucleus is shattered and the neutron
is liberated.  Nevertheless, the heavy water detector observed the CC reaction
above and the ES reaction rates by 2001, but had to wait until 2002 to
observe the NC.  The latter required that ultra-pure common salt (NaCl)
be introduced into the heavy water so that the neutrons produced in the NC
reactions could be absorbed by the Cl nuclei and then produce a characteristic
8.6 MeV gamma ray signal. Otherwise neutrons were detected by characteristic 6.25
MeV gamma ray when they get absorbed by deuterium in heavy water. The neutron
absorption probability in heavy water is about 25\% which increased to 85\% by
addition of ultra-pure NaCl. This was achieved and the results subsequently 
reported in Ref.~\cite{nc}.  The final numbers quoted therein translate to 2/3 
of the electron-type neutrinos oscillating into muon and tau type flavors.
Furthermore, the observation of a non-vanishing day night effect shows
that there may be some regeneration of the electron type neutrino flux
in the passage of the neutrinos from the sun through the earth.  Such an
effect, known as the Mikheyev-Smirnov-Wolfenstein effect, has been studied
in the past and is now likely to be constrained quite effectively, or 
alternative vacuum oscillation scenarios are likely to be constrained
as well.  For a recent discussion of the impact of the SNO measurements
on theoretical scenarios, we refer to ref.~\cite{bahcall}.\\\\
It must be mentioned again that the advantages of the heavy water also leads
to the possibility of large backgrounds.  In fact, the SNO experiment is
located deep underground in nickel mines in Canada, and the heavy water which
is stored in a large acrylic container is also surrounded by jackets containing
normal water in order to absorb radiation from the surrounding rock and also
from cosmic ray sources which could easily generate spurious signals which
serve as a background. \\\\
In conclusion, we note that the remarkable experiment at SNO based on 
the deep insights of Chen has resolved the
solar neutrino problem in favor of a solution arising from neutrino
oscillations, rather than from unknown inadequacies of the standard solar
model.  The SNO collaboration is expected to improve its statistics and
bring down uncertainties in their measurements and will pave the way to
confirming and constraining theoretical scenarios which account for neutrino
oscillations. 

The article so far has been identical to Version 1.  

{\bf Errata to Version 1:}
The published results of SNO on the evidence for neutrino oscillations
relies on the observation of the gamma radiation from neutron capture on
deuterium, and not from data taken after salt addition, as incorrectly
stated in our article.  The radiochemical measurements used
tetrachloroethylene ($C_2\, Cl_4$, also called perchloroethylene), 
not carbon tetrachloride ($C\, Cl_4$) as stated.  
It may also be noted that the ES reactions also involve NC and not
just CC events.  We thank the SNO collaboration and
M. V. N. Murthy for pointing out these errors.

\end{small}
\end{document}